\documentclass[twocolumn,aps,floats,floatfix, superscriptaddress,prl]{revtex4-1}
\usepackage{graphicx,amssymb}
\usepackage{epsfig}
\usepackage{mathtools}
\usepackage{soul}

\DeclarePairedDelimiter\ket{\lvert}{\rangle}

\begin{document}
\title{Spin-orbit coupled two-electron Fermi gases of ytterbium atoms}

\author{Bo Song}
\affiliation{Department of Physics, The Hong Kong University of Science and Technology, \\Clear Water Bay, Kowloon, Hong Kong, China}

\author{Chengdong He}
\affiliation{Department of Physics, The Hong Kong University of Science and Technology,  \\Clear Water Bay, Kowloon, Hong Kong, China}

\author{Shanchao Zhang}
\affiliation{Department of Physics, The Hong Kong University of Science and Technology,  \\Clear Water Bay, Kowloon, Hong Kong, China}

\author{Elnur Hajiyev}
\affiliation{Department of Physics, The Hong Kong University of Science and Technology, \\Clear Water Bay, Kowloon, Hong Kong, China}

\author{Wei Huang}
\altaffiliation{Present address: Guangdong Provincial Key Laboratory of Quantum Engineering and Quantum Materials, School of Physics and Telecommunication Engineering, South China Normal University, Guangzhou, China.}
\affiliation{Department of Physics, The Hong Kong University of Science and Technology,  \\Clear Water Bay, Kowloon, Hong Kong, China}

\author{Xiong-Jun Liu}
\affiliation{International Center for Quantum Materials, School of Physics, Peking University, Beijing, China}

\author{Gyu-Boong Jo}
\email{gbjo@ust.hk}
\affiliation{Department of Physics, The Hong Kong University of Science and Technology, \\Clear Water Bay, Kowloon, Hong Kong, China}

\begin{abstract}
We demonstrate all-optical implementation of spin-orbit coupling (SOC) in a two-electron Fermi gas of $^{173}$Yb atoms by coupling two hyperfine ground states with a narrow optical transition. Due to the SU($N$) symmetry of the $^1$S$_0$ ground-state manifold which is insensitive to external magnetic fields, an optical AC Stark effect is applied to split the ground spin states, which exhibits a high stability compared with experiments on alkali and lanthanide atoms, and separate out an effective spin-1/2 subspace from other hyperfine levels for the realization of SOC. The dephasing spin dynamics when a momentum-dependent spin-orbit gap being suddenly opened and the asymmetric momentum distribution of the spin-orbit coupled Fermi gas are observed as a hallmark of SOC. The realization of all-optical SOC for ytterbium fermions should offer a new route to a long-lived spin-orbit coupled Fermi gas and greatly expand our capability in studying novel spin-orbit physics with alkaline-earth-like atoms.
\end{abstract}
\maketitle




Ultracold atoms are fascinating for the study of synthetic quantum system which is direct analogy to real electronic material~\cite{Bloch:2008gl}. One of the notable examples is the implementation of synthetic gauge field and spin-orbit coupling (SOC) engineered with the atom-light interaction at will~\cite{Dalibard:2011gg,Zhai:2015hg}. In particular, SOC links a particle's spin with its momentum, which is not only essential in  novel quantum phenomena, such as spintronic effect~\cite{Nagaosa:2010js} and exotic topological states of quantum matter~\cite{Hasan:2010ku,Qi:2011wt}, but also provides an unprecedented quantum system such as spin-half spin-orbit coupled bosons without analogy in condensed-matter~\cite{Lin:2011hn}. Various types of SOCs can be generated in ultracold atoms where the relevant parameters are tunable by changing the laser fields~\cite{Higbie:2002gb,Liu:2009hj,Spielman:2009ej} or the magnetic field~\cite{Luo:2016iw}. So far, the SOCs along the one direction have been created in bosonic alkali~\cite{Lin:2011hn,Jimenez-Garcia2015,Williams:2012gs,Zhang2012,Qu2013,Hamner2014,Ji2014,Olson2014,Ji2015},  fermionic alkali atoms~\cite{Wang2012,2012PhRvL.109i5302C,Williams2013,Fu2014}, and very recently in fermionic lanthanide atoms~\cite{NathanieQ2016}. Besides the 1D SOC, the two-dimensional synthetic SOCs have been also demonstrated both in the bosonic~\cite{2015arXiv151108170W} and fermionic alkali atoms~\cite{Huang2016}. 

In alkali atoms, two different internal states are coupled through the Raman transition transferring momentum to the atoms~\cite{Dalibard:2011gg,Zhai:2015hg}. However those processes inevitably suffer from heating effect caused by spontaneous emission due to the small fine-structure splitting of the excited level, which could limit the ability to observe interacting many-body phenomena that needs long timescales. Recently to avoid such heating, the specific atomic species with the large ground-state angular momentum such as $^{161}$Dy have been considered~\cite{Cui2013,Wall2016} or the external orbital states, representing pseudo-spins, in optical superlattices have been used to generate SOC~\cite{Atala:2014gf,Li:2016vc}. 

Here, we expand our capability in exploring a novel SOC physics by implementing SOC with a narrow optical transition in a non-alkali Fermi gas of ytterbium atoms. With a momentum-dependent spin-orbit gap being suddenly opened by switching on the Raman transition, the dephasing of spin dynamics is observed, as a consequence of the momentum-dependent Rabi oscillation. Moreover, the momentum asymmetry of the spin-orbit coupled Fermi gas is  examined after projection onto the bare spin state and the corresponding momentum distribution is measured for different values of two-photon detuning. In contrast to spin-orbit-coupled alkali or lanthanide fermionic atoms, here the two-photon detuning is fully controlled by the light-induced AC Stark shift for $^{173}$Yb atoms. The present technique exhibits several potential advantages in the study of SOC compared with alkali and lanthanide experiments, including the precise control of spin splitting without necessitating extreme magnetic field stability to reduce heating, and the creation of spin textures in momentum and spatial dimension by locally modulating the two-photon detuning. These advances demonstrated in this work, combined by all-optical engineering of SOC, can expand variety of spin-orbit coupled Fermi gases. 

Indeed recent development of alkaline-earth-like atoms has led to an increasing potential in utilizing such atoms not only for the study of many-body physics~\cite{2011arXiv1106.5712D,Cazalilla:2014ut,Pagano:2014hy,Taie:2012vp} but also for the realization of synthetic gauge field in a Fermi gas~\cite{Mancini2015,2016NatPh..12..296N}. A narrow optical transition of such atoms and their large fine-structure splitting, for example, should allow to alleviate light-induced heating when SOC is generated. Moreover, the spin splitting by the AC Stark shift in our present study has a stability over one order of magnitude better than the typical magnetic field stability in alkali or lanthanide experiments. This further avoids the heating induced by fluctuations of the spin splitting. Finally, the alkaline-earth-like atom with large nuclear spins offers a new possibility to realize spin-orbit coupled high-spin fermions~\cite{Wu:2003es,Gorshkov:2010hw,Zhai:2015hg,Cai:2013ke}, intriguing magnetic crystals~\cite{Barbarino:2015fi} and atomic simulator of U($N$) and SU($N$) lattice gauge theories~\cite{Banerjee:2013fl}. 



\begin{figure}[tbp]
\resizebox{7.7cm}{!}{\includegraphics{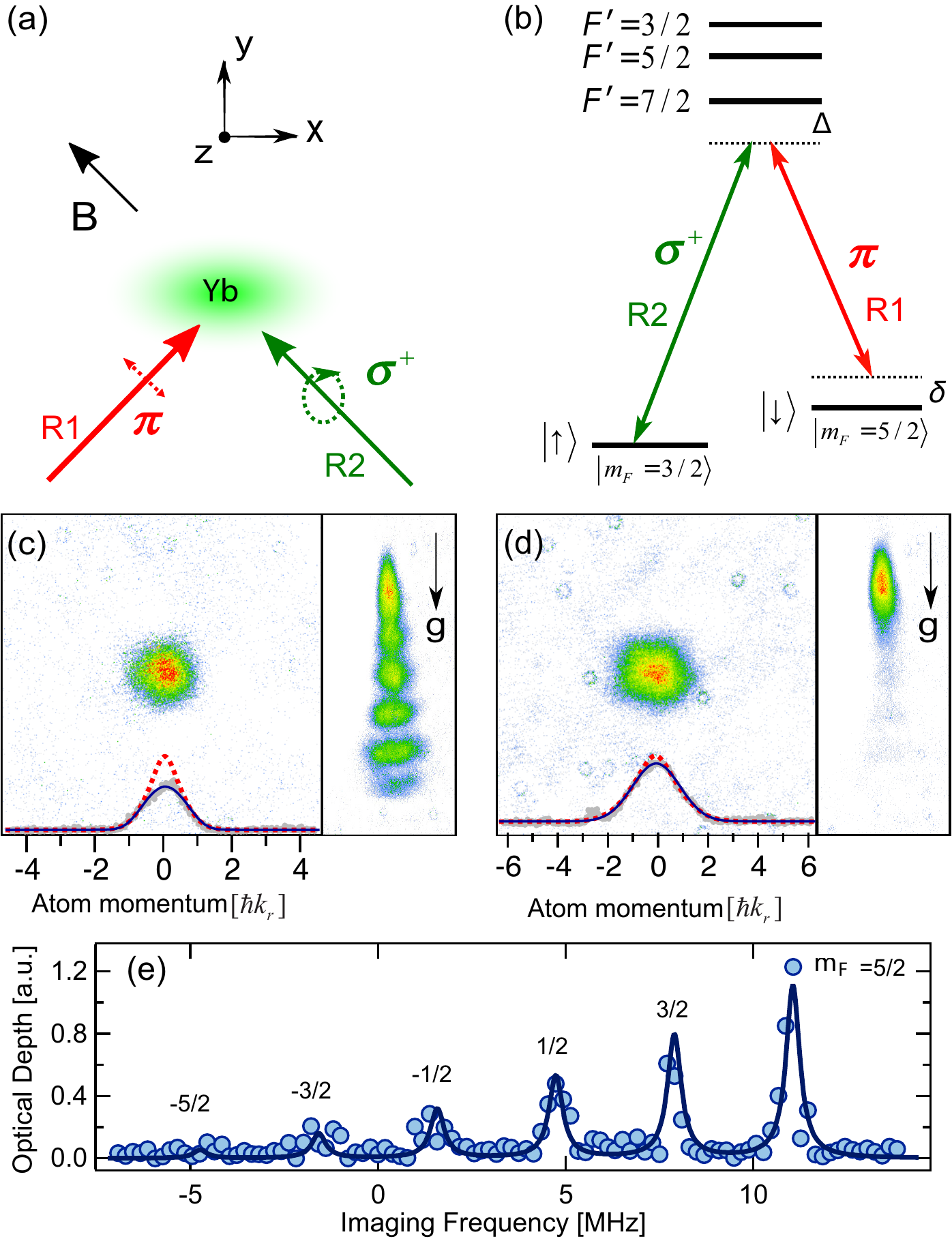}}
\caption{ (color online) Schematic diagram of the Raman transition configuration. (a) A single-component fermi gas of $^{173}$Yb atoms is prepared with bias magnetic field along the $(-\hat{x}+\hat{y})$ direction and exposed to a pair of two 556~nm Raman beams (linearly and circularly polarized respectively)  at 45$^{\circ}$ relative to $\hat{y}$ axis. The direction of the gravity is along $(\hat{x}-\hat{y})$ direction. (b) A pair of Raman beams couples two ground-states hyperfine level with the single-photon detuning  $\Delta=1$~GHz from the excited state $F'$=$\frac{7}{2}$. (c) A six-component Fermi gas of $^{173}$Yb atoms with $N=5.0\times 10^4$ is obtained at $T/T_F\leq 0.27$. A time-of-flight image is obtained after 11~ms free expansion using the 399~nm ${}^1$S$_0 - {}^1$P$_1$ transition. Each spin state $\ket{m_F}$ is spatially separated by the OSG light followed by 399~nm absorption imaging. (d) With optical pumping during the evaporative cooling, a single-component Fermi gas in $\ket{m_F=\frac{5}{2}}$ is cooled down to $T/T_F\simeq 0.5$. (e) Using the ${}^1$S$_0 - {}^3$P$_1$ intercombination line, $m_F$-states are resolved. The spectrum of $^{173}$Yb atoms is obtained by a $\sigma^{+}$-polarized light in the 5~G bias field along the imaging axis. The black solid curve indicates the expected relative line strength of the transition.}
\label{Fig1}
\end{figure}



\paragraph{All-optical implementation of SOC in an ytterbium gas}
We generate spin-orbit coupling in a Fermi gas of $^{173}$Yb atoms by the use of  the intercombination $\lambda_0$=556~nm transition connecting two hyperfine levels with the Raman transition~\cite{Wang2012,2012PhRvL.109i5302C}. A pair of Raman coupling beams {\color{red} with the diameter of 70 $\mu$m and the power of 100~$\mu$W}, intersecting at  $\theta=$90$^{\circ}$, are red-detuned by $\sim$1~GHz from the ${}^1$S$_0 (F = \frac{5}{2}) \leftrightarrow {}^3$P$_1 (F' = \frac{7}{2})$ narrow linewidth transition. The $F=$$\frac{5}{2}$ ground-state manifold has six hyperfine levels with magnetic sublevels $\lvert m_F\rvert \leq$$\frac{5}{2}$ that are typically degenerate due to their small magnetic moment which entirely stems from the nuclear spin.


Raman process couples two hyperfine levels $\ket{F=\frac{5}{2},m_F=\frac{3}{2}}$ and $\ket{F=\frac{5}{2},m_F=\frac{5}{2}}$ in the $^1$S$_0$ ground state, labeled as spin-up $\ket{\uparrow}$ and spin-down $\ket{\downarrow}$ respectively, by imparting momentum 2$\hbar k_r$ to atoms where the recoil momentum is $k_r=k_0 \sin(\theta/2)$ for $k_0=2\pi/\lambda_0$. Each of Raman beams induces the spin-dependent AC Stark shift splitting the magnetic sublevels. We achieve the Hamiltonian that describes effective 1D SOC, with equal strengths of Rashba and Dresselhaus SOC, along the $\hat{x}$-direction as the previous work~\cite{Wang2012,2012PhRvL.109i5302C,Lin:2011hn,NathanieQ2016}: $ H_{SOC}=\frac{1}{2m}(p_x + k_r \sigma_z)^2 + \frac{\Omega_R}{2}\sigma_x+\frac{\delta}{2}\sigma_z$ where $\sigma_i$ are Pauli matrices, $p_x$ is the quasi-momentum of atoms along $\hat{x}$-direction, $m$ is the mass of ytterbium atom, $\Omega_R$ is the Rabi frequency of Raman coupling and $\delta$ is the two-photon detuning. In contrast to alkali or lanthanide atoms, we tune the two-photon detuning $\delta$ using spin-dependent light shift induced by  Raman beams. We independently calibrate spin-dependent level shift from different lights by monitoring the two-photon Rabi oscillation induced by a pair of co-propagating Raman beams (see appendix). 





\paragraph{Preparation of the ytterbium gas}
We first prepare an ultracold Fermi gas of $^{173}$Yb atoms by loading atoms, precooled by the intercombination magneto-optical trap (MOT), into the a 1064~nm crossed optical dipole trap (ODT) for forced evaporation~\cite{Fukuhara:2007wg}. After the final stage of the optical evaporative cooling, we achieve a six-component degenerate Fermi gas of $\sim$5.0$\times 10^4$ atoms at $T/T_F\leq$~0.27 where $T_F$ is the Fermi temperature of the trapped atom with the trapping frequency of $\overline{\omega}=(\omega_x\omega_y\omega_z)^{\frac{1}{3}}$=$2\pi\times$130~Hz. 

A single-component degenerate Fermi gas is then obtained utilizing the similar optical evaporative cooling process with different initial spin configuration. We optically pump more than 70$\%$ atoms into the $\ket{m_F=\frac{5}{2}}$ state at the beginning of the evaporative cooling using the nearly resonant 556~nm light with $\sigma^+$ polarization. Note that we keep a small fraction of $\ket{m_F=\frac{3}{2}}$ state atoms until the end of the evaporation in order to enhance the evaporative cooling. Finally, we prepare a Fermi gas of $N=1.2\times 10^4$ atoms in $\ket{m_F=\frac{5}{2}}$ state at $T/T_F\simeq 0.5$ 


\paragraph{Spin-resolved detection}
For ytterbium isotopes, $m_F$-resolved absorption imaging with the broad 399~nm ${}^1$S${}_0-^1$P$_1$ transition is not possible at low bias magnetic field as the Zeeman splitting of the excited $\ket{m_{F'}}$ state is not large enough compared to the natural linewidth of the transition. Here we instead use the narrow $^1$S$_0-^3$P$_1$ intercombination line for spin-resolved blast at the bias field of 5~G where the ground-state manifold is degenerate with SU(6) symmetry~\cite{Stellmer:2011iv}. As described in Fig.~\ref{Fig1} (e), each $\ket{m_F}$ state is spectroscopically resolved at different imaging frequency of 556~nm light following the line strength of the corresponding transition. During the time-of-flight expansion, the unwanted spin states are selectively removed with the blast light~\cite{2015arXiv150202495M}, followed by the 399~nm absorption imaging.


We also use the complimentary method to resolve different nuclear spin states by  optical Stern-Gerlach (OSG) effect~\cite{Stellmer:2011iv,Taie:2010uo}.  A non-resonant circularly polarized laser beam with the beam waist of 140~$\mu m$ is briefly switched on after the time-of-flight expansion starts. The spin-dependent force vertically splits the atomic cloud 
into six parts corresponding each $\ket{m_F}$ state as shown in Fig.~\ref{Fig1} (c) and (d) inset.


\begin{figure}[tbp]
\resizebox{8.6cm}{!}{\includegraphics{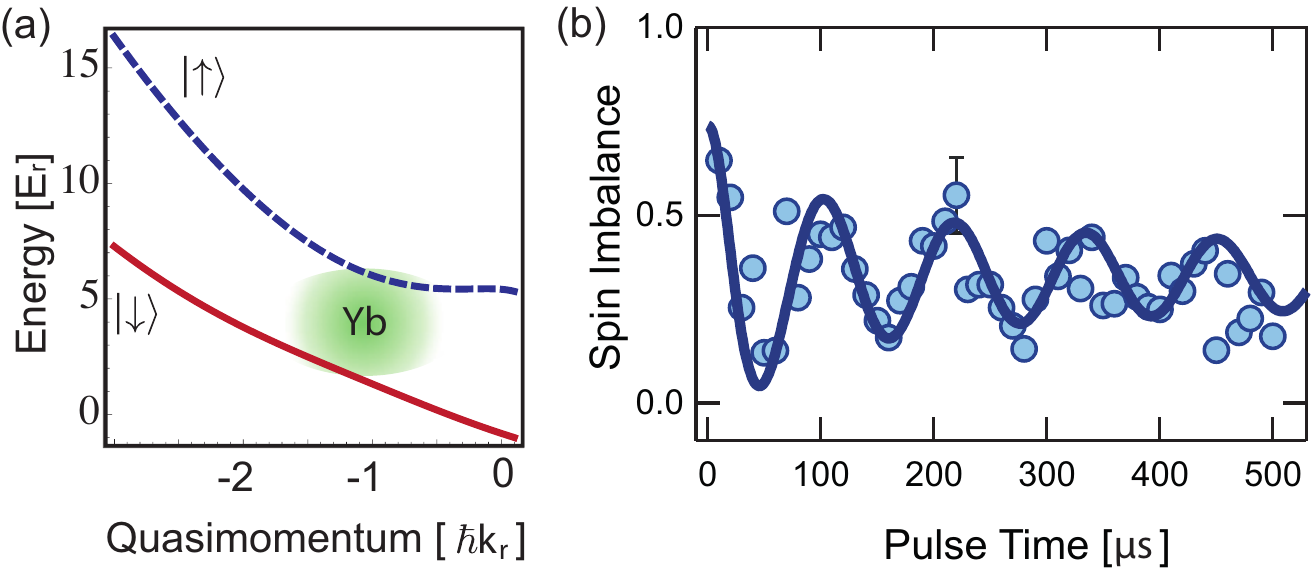}}
\caption{ (color online)  Spin dynamics with spin-orbit coupling. The two-photon Rabi oscailltion is monitored with the two-photon detuning of $\delta$=$-4.0(3)E_r$  and the Rabi frequency $\Omega_R=4.8(3) E_r$. (a) The single-particle energy dispersion of dressed-states. (b) The spin imbalance, defined as $(N_{\ket{\downarrow}}-N_{\ket{\uparrow}})/(N_{\ket{\downarrow}}+N_{\ket{\uparrow}})$ for the atom number $N_{\ket{\uparrow}}$ and $N_{\ket{\downarrow}}$, is measured via 399~nm imaging together with the OSG light. The predicted spin dynamics for non-interacting fermions (blue solid curve) is consistent with the measurement (blue circle). The data points represent the average of ten measurements.} 
\label{Fig2}
\end{figure}

\paragraph{Spin dynamics with spin-orbit coupling}
When two spin states $\ket{\uparrow}$  and $\ket{\downarrow}$ are coupled through the two-photon Raman transition with momentum transfer of $2\hbar{k}_r \mathbf{\hat{e}_x}$, the energy gap between the dressed-states $\ket{\uparrow^{'},\mathbf{p}_x=\mathbf{k}_x-\hbar k_r\mathbf{\hat{e}_x}}$ and $\ket{\downarrow^{'},\mathbf{p}_x=\mathbf{k}_x+\hbar k_r\mathbf{\hat{e}_x}}$  is quasi-momentum dependent where $\mathbf{p}_x$ denotes the quasi-momentum as shown in Fig.~\ref{Fig2} (a). When a non-uniform spin-orbit gap being opened, fermions with different quasi-momenta undergo spin precession in different ways and the total spin oscillation will be damped out~\cite{Wang2012,2012PhRvL.109i5302C,NathanieQ2016}. In this measurement, a brief pulse of Raman-coupling beams is switched on and subsequently turned off while the atoms remain in the crossed ODT.
The bias magnetic field is then properly rotated to the imaging axis along the $(\hat{z}+\hat{y})$ direction within 200~ms during which the nuclear spin follows the bias field adiabatically. We experimentally confirm that the spin distribution is not changed during that process owing to SU($N$) symmetry. Each spin state is then spatially separated by applying the OSG light followed by 4~ms time-of-flight free expansion before the 399~nm absorption imaging. {\color{red}The oscillation of spin distribution is dephased as a function of pulse time which is the hallmark of the spin-orbit coupling as described in Fig.~\ref{Fig2}. 
The two-photon Rabi frequency $\Omega_R$ coupling $\ket{\uparrow}$ and $\ket{\downarrow}$ is then measured by fitting the oscillation of the spin population with the momentum-dependent Rabi oscillation for non-interacting fermions~\cite{Wang2012,NathanieQ2016} as shown in Fig.~\ref{Fig2} (b). }


\begin{figure}[tbp]
\resizebox{8.2cm}{!}{\includegraphics{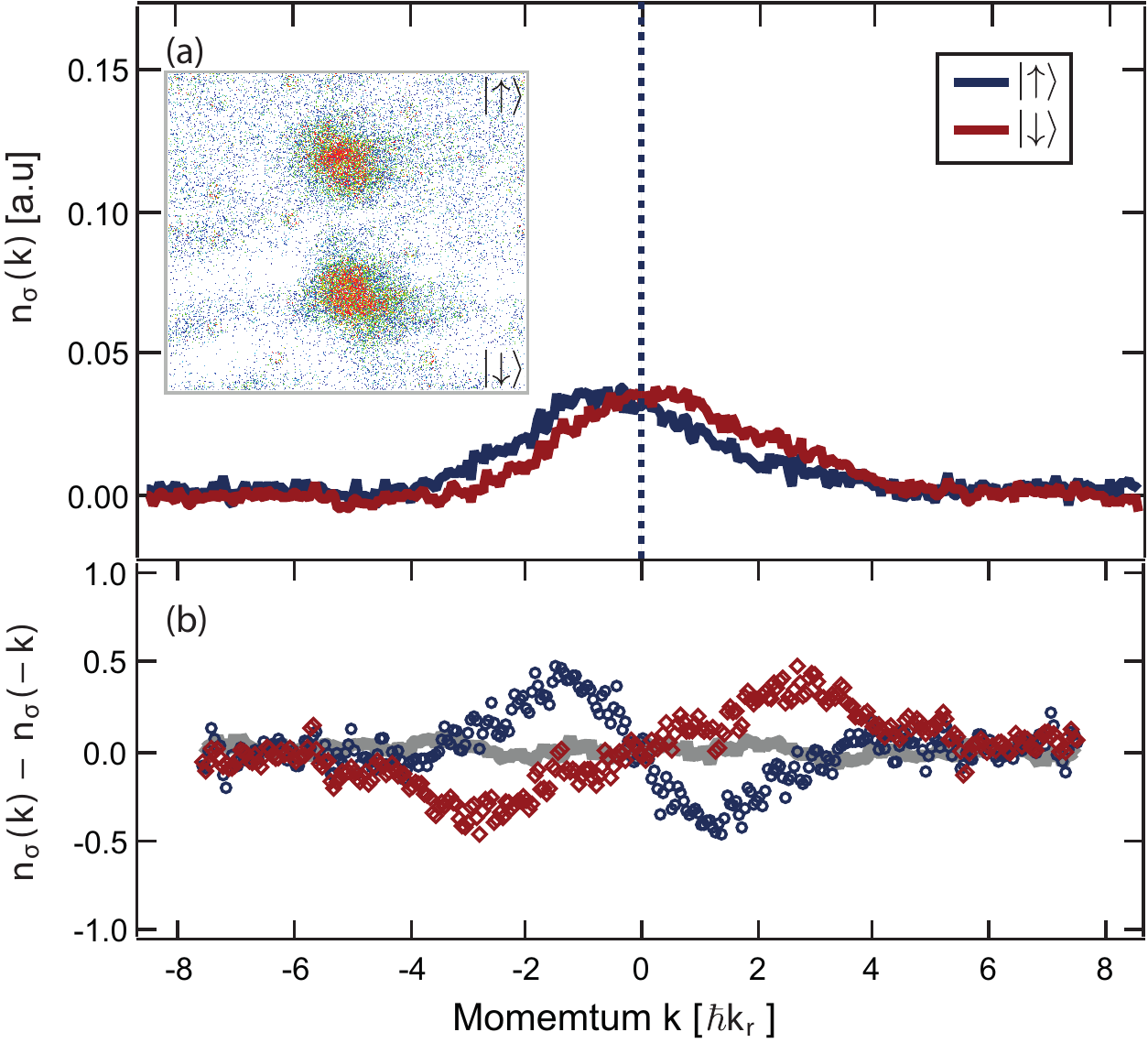}}
\caption{ (color online) Momentum distribution of a spin-orbit coupled Fermi gas of $^{173}$Yb  atoms. The Raman coupling with $\Omega = 1.6(3) E_R$ and $\delta=0.1(3)$ is abruptly switched off for a gas of $N$=$9\times 10^3$ atoms followed by 399~nm absorption imaging as shown in the inset. (a) Atomic clouds after time-of-flight expansion for spin $\sigma=\{\uparrow,\downarrow\}$ shift in the opposite direction due to SOC. (b) The integrated momentum reveals asymmetric feature for $\ket{\uparrow}$ (blue circle) and $\ket{\downarrow}$ (red diamond) while the momentum distribution is symmetric without SOC (grey curve).}
\label{Fig3}
\end{figure}

\paragraph{Adiabatic loading into the dressed-state}
In order to adiabatically load atoms into the equilibrium state of the lowest-energy dressed state~\cite{CohenTannoudji:1992wp}, we start with a spin-polarized $\ket{m_F=\frac{5}{2},\downarrow}$ Fermi gas with $T/T_F\simeq$0.5 confined in a crossed dipole trap and increase the Raman-coupling beam power with a 7~ms exponential ramp to the final values keeping the two-photon detuning $\delta\simeq$~10$E_r$ where the recoil energy is $E_r=\hbar^2 k_r^2/2m$=1.87~kHz. Due to the large detuning, the atoms in the bare $\ket{\downarrow}$ state are adiabatically loaded into the dressed spin state $\ket{\downarrow^{'}}$, by which we determine the momentum of $k_x$=~0 in the time-of-flight expansion. Subsequently, the two-photon detuning $\delta$ is exponentially ramped to the final value of detuning within 23~ms.

\begin{figure}[tbp]
\resizebox{8.7cm}{!}{\includegraphics{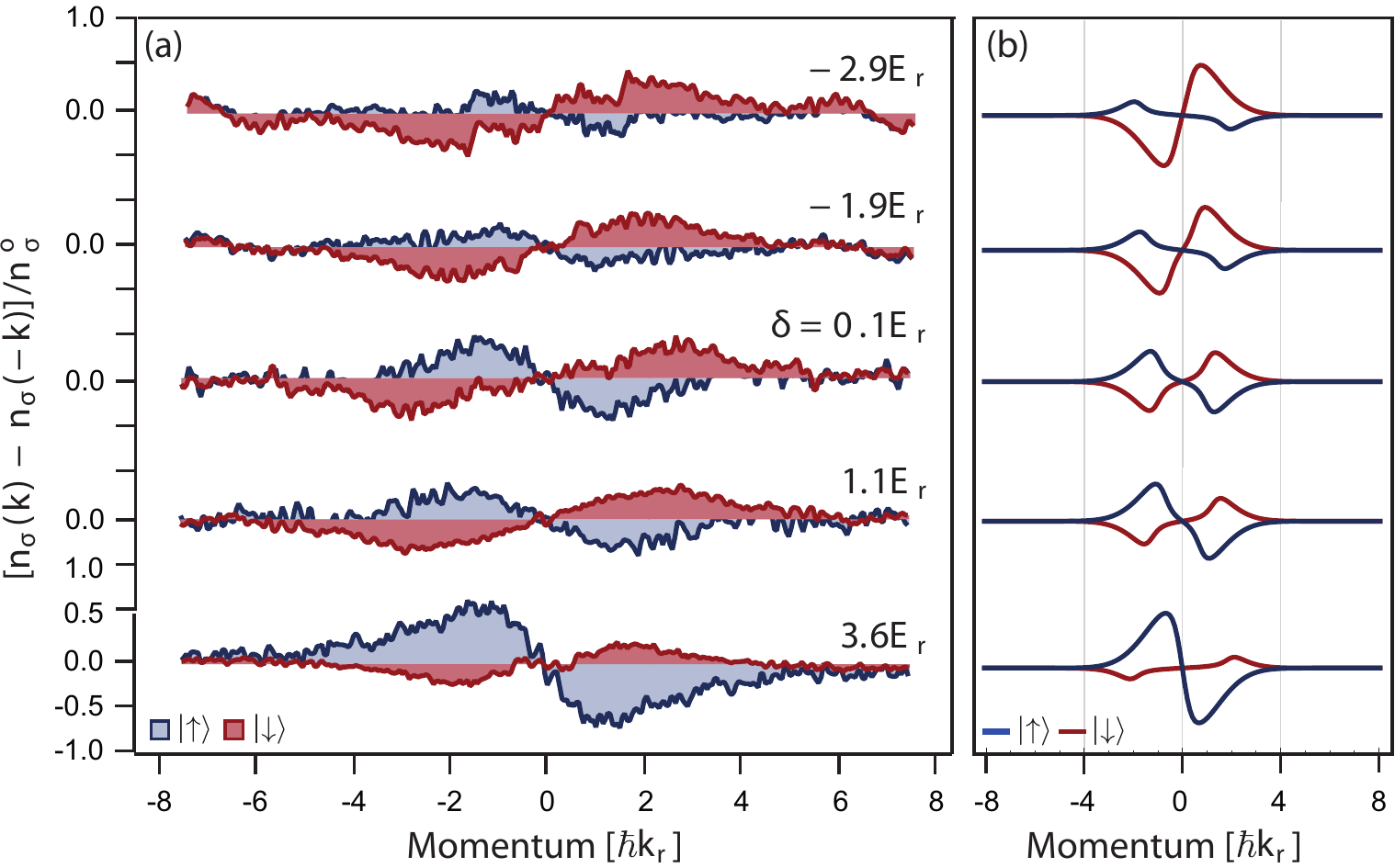}}
\caption{ (color online) Integrated momentum asymmetry along $\hat{x}$-direction for different values of two-photon detuning. The momentum asymmetry of the spin-orbit coupled Fermi gas is characterized by $[n_{\sigma}(k)-n_{\sigma}(-k)]/n_{\sigma}^o$ where $n_{\sigma}^o=max[n_{\sigma}(k)]$ denotes the maximum peak density of the momentum distribution in $k$-space for $\sigma=\{\uparrow,\downarrow\}$. The two-photon detuning takes the different values of  $\delta=\{-2.9,-1.9,0.1,1.1,3.6\} \times E_r$ from top to bottom. (b) The theoretical momentum asymmetry is shown for different values of detuning. As $\delta>0$ becomes large where most of atoms are loaded into the $\ket{\downarrow'}$ state, the momentum asymmetry of $\ket{\downarrow}$ disappears after projection onto the bare state while $\ket{\uparrow}$ atoms show prominent momentum asymmetry due to spin-orbit coupling.  }
\label{Fig4}
\end{figure}

After adiabatic loading into the dressed state with $\Omega_R=$1.6(3)$E_r$ and $\delta=$0.1(3)$E_r$, we abruptly switch off the Raman beams and the crossed ODT simultaneously - thus projecting the dressed states onto the bare spin $\ket{\uparrow,\downarrow}$ and the real momentum $k_x$ - followed by the 399~nm absorption imaging after 7~ms time-of-flight expansion with spin-selective blast 556~nm light. When the dressed energy band is symmetric for $\ket{\uparrow}$ and $\ket{\downarrow}$ at $\delta=$0.1(3)$E_r$, the SOC, breaking spatial reflection symmetry, introduces asymmetric momentum distribution for bare spin states~\cite{Wang2012,NathanieQ2016}, as highlighted in Fig.~\ref{Fig3} where the integrated momentum distribution (blue and red curves) and $n_\sigma(k)-n_\sigma(-k)$ are plotted for spin $\sigma=\{\uparrow,\downarrow\}$. We note that such momentum asymmetry disappears if the SOC is switched off as shown by a grey curve. The $1/e$ lifetime for typical Raman coupling setting with a one-photon detuning $\Delta\sim$ 1.0~GHz is about 70~ms that is much longer than the alkali system. We note that if a larger one-photon detuning is applied for the $^{173}$Yb atoms an even longer lifetime is expected than Dy~\cite{2016arXiv160503211B} or 2D spin-orbit-coupled Rb atoms~\cite{2015arXiv151108170W}. In the present work, the single-photon detuning is limited by the available laser power of the narrow optical transition at 556~nm light. Finally for the range of detuning $\delta \gtrsim-1.0E_r$, where the $\ket{m_F=\frac{5}{2},\frac{3}{2}}$ state is detuned by at least $\sim$2$E_r$ from the $\ket{m_F=\frac{1}{2}}$ state, the system is effectively described by spin-$\frac{1}{2}$ SOC hamiltonian keeping the fractional population in the unwanted state $\ket{m_F=\frac{1}{2}}$ smaller than 0.1. In Fig.~\ref{Fig4}, we cross-validate this behaviour by comparing the momentum asymmetry to the predicted distribution calculated by considering the eigenstate of the single-particle Hamiltonian $H_{SOC}$ over the momentum distribution of the Fermi gas~\cite{sup}.

\paragraph{Discussion and Conclusion}
We have demonstrated all-optical SOC in a degenerate Fermi gas of $^{173}$Yb atoms by using the Raman transition. The momentum distribution of the spin-orbit coupled Fermi gas exhibits strong asymmetry as a hallmark of SOC. The spin splitting is induced by optical AC Stark shift, which can be precisely controllable by tuning light intensity. This avoids applying magnetic field as in the experiments with alkali or lanthanide atoms, which suffers from magnetic field fluctuations. The stability of the two-photon detuning in this work is equivalent to the field stability of a few tens of $\mu G$ in the alkali system which is one order of magnitude better than current experiments. Furthermore, based on the present experiment a rich configuration, including spatially dependent two-photon detuning can be easily created by spatial dependent light strengths. Such configuration may create a spin texture in both momentum and position space. Finally, without the limit of fine structure splitting for excited levels, a longer lifetime can be obtained for SOC with ytterbium atoms when a large one-photon detuning is considered for the optical transitions~\cite{Cui:2013ki}. Being of many novel features in the ground manifold of hyperfine states, including the SU($N$) symmetry, the ytterbium fermions with all-optical SOC would open new possibilities to explore interesting new quantum physics.

\section{Acknowledement}
G.-B. J. acknowledges the generous support from the Hong Kong Research Grants Council and the Croucher Foundation through ECS26300014, GRF16300215, GRF16311516 and the Croucher Innovation grants respectively. X.-J. L. also thanks the support from MOST (Grant No. 2016YFA0301600) and NSFC (No. 11574008).


\begin{thebibliography}{48}%
\makeatletter
\providecommand \@ifxundefined [1]{%
 \@ifx{#1\undefined}
}%
\providecommand \@ifnum [1]{%
 \ifnum #1\expandafter \@firstoftwo
 \else \expandafter \@secondoftwo
 \fi
}%
\providecommand \@ifx [1]{%
 \ifx #1\expandafter \@firstoftwo
 \else \expandafter \@secondoftwo
 \fi
}%
\providecommand \natexlab [1]{#1}%
\providecommand \enquote  [1]{``#1''}%
\providecommand \bibnamefont  [1]{#1}%
\providecommand \bibfnamefont [1]{#1}%
\providecommand \citenamefont [1]{#1}%
\providecommand \href@noop [0]{\@secondoftwo}%
\providecommand \href [0]{\begingroup \@sanitize@url \@href}%
\providecommand \@href[1]{\@@startlink{#1}\@@href}%
\providecommand \@@href[1]{\endgroup#1\@@endlink}%
\providecommand \@sanitize@url [0]{\catcode `\\12\catcode `\$12\catcode
  `\&12\catcode `\#12\catcode `\^12\catcode `\_12\catcode `\%12\relax}%
\providecommand \@@startlink[1]{}%
\providecommand \@@endlink[0]{}%
\providecommand \url  [0]{\begingroup\@sanitize@url \@url }%
\providecommand \@url [1]{\endgroup\@href {#1}{\urlprefix }}%
\providecommand \urlprefix  [0]{URL }%
\providecommand \Eprint [0]{\href }%
\providecommand \doibase [0]{http://dx.doi.org/}%
\providecommand \selectlanguage [0]{\@gobble}%
\providecommand \bibinfo  [0]{\@secondoftwo}%
\providecommand \bibfield  [0]{\@secondoftwo}%
\providecommand \translation [1]{[#1]}%
\providecommand \BibitemOpen [0]{}%
\providecommand \bibitemStop [0]{}%
\providecommand \bibitemNoStop [0]{.\EOS\space}%
\providecommand \EOS [0]{\spacefactor3000\relax}%
\providecommand \BibitemShut  [1]{\csname bibitem#1\endcsname}%
\let\auto@bib@innerbib\@empty
\bibitem [{\citenamefont {Bloch}\ and\ \citenamefont
  {Zwerger}(2008)}]{Bloch:2008gl}%
  \BibitemOpen
  \bibfield  {author} {\bibinfo {author} {\bibfnamefont {I.}~\bibnamefont
  {Bloch}}\ and\ \bibinfo {author} {\bibfnamefont {W.}~\bibnamefont
  {Zwerger}},\ }\href@noop {} {\bibfield  {journal} {\bibinfo  {journal}
  {Reviews of Modern Physics}\ }\textbf {\bibinfo {volume} {80}},\ \bibinfo
  {pages} {885} (\bibinfo {year} {2008})}\BibitemShut {NoStop}%
\bibitem [{\citenamefont {Dalibard}\ \emph {et~al.}(2011)\citenamefont
  {Dalibard}, \citenamefont {Gerbier}, \citenamefont {Juzeli{\=u}nas},\ and\
  \citenamefont {Oehberg}}]{Dalibard:2011gg}%
  \BibitemOpen
  \bibfield  {author} {\bibinfo {author} {\bibfnamefont {J.}~\bibnamefont
  {Dalibard}}, \bibinfo {author} {\bibfnamefont {F.}~\bibnamefont {Gerbier}},
  \bibinfo {author} {\bibfnamefont {G.}~\bibnamefont {Juzeli{\=u}nas}}, \ and\
  \bibinfo {author} {\bibfnamefont {P.}~\bibnamefont {Oehberg}},\ }\href@noop
  {} {\bibfield  {journal} {\bibinfo  {journal} {Reviews of Modern Physics}\
  }\textbf {\bibinfo {volume} {83}},\  (\bibinfo {year} {2011})}\BibitemShut
  {NoStop}%
\bibitem [{\citenamefont {Zhai}(2015)}]{Zhai:2015hg}%
  \BibitemOpen
  \bibfield  {author} {\bibinfo {author} {\bibfnamefont {H.}~\bibnamefont
  {Zhai}},\ }\href@noop {} {\bibfield  {journal} {\bibinfo  {journal} {Reports
  on progress in physics. Physical Society (Great Britain)}\ }\textbf {\bibinfo
  {volume} {78}},\ \bibinfo {pages} {026001} (\bibinfo {year}
  {2015})}\BibitemShut {NoStop}%
\bibitem [{\citenamefont {Nagaosa}\ \emph {et~al.}(2010)\citenamefont
  {Nagaosa}, \citenamefont {Sinova}, \citenamefont {Onoda}, \citenamefont
  {MacDonald},\ and\ \citenamefont {Ong}}]{Nagaosa:2010js}%
  \BibitemOpen
  \bibfield  {author} {\bibinfo {author} {\bibfnamefont {N.}~\bibnamefont
  {Nagaosa}}, \bibinfo {author} {\bibfnamefont {J.}~\bibnamefont {Sinova}},
  \bibinfo {author} {\bibfnamefont {S.}~\bibnamefont {Onoda}}, \bibinfo
  {author} {\bibfnamefont {A.~H.}\ \bibnamefont {MacDonald}}, \ and\ \bibinfo
  {author} {\bibfnamefont {N.~P.}\ \bibnamefont {Ong}},\ }\href@noop {}
  {\bibfield  {journal} {\bibinfo  {journal} {Reviews of Modern Physics}\
  }\textbf {\bibinfo {volume} {82}},\ \bibinfo {pages} {1539} (\bibinfo {year}
  {2010})}\BibitemShut {NoStop}%
\bibitem [{\citenamefont {Hasan}\ and\ \citenamefont
  {Kane}(2010)}]{Hasan:2010ku}%
  \BibitemOpen
  \bibfield  {author} {\bibinfo {author} {\bibfnamefont {M.}~\bibnamefont
  {Hasan}}\ and\ \bibinfo {author} {\bibfnamefont {C.}~\bibnamefont {Kane}},\
  }\href@noop {} {\bibfield  {journal} {\bibinfo  {journal} {Reviews of Modern
  Physics}\ }\textbf {\bibinfo {volume} {82}},\ \bibinfo {pages} {3045}
  (\bibinfo {year} {2010})}\BibitemShut {NoStop}%
\bibitem [{\citenamefont {Qi}\ and\ \citenamefont {Zhang}(2011)}]{Qi:2011wt}%
  \BibitemOpen
  \bibfield  {author} {\bibinfo {author} {\bibfnamefont {X.~L.}\ \bibnamefont
  {Qi}}\ and\ \bibinfo {author} {\bibfnamefont {S.~C.}\ \bibnamefont {Zhang}},\
  }\href@noop {} {\bibfield  {journal} {\bibinfo  {journal} {Reviews of Modern
  Physics}\ } (\bibinfo {year} {2011})}\BibitemShut {NoStop}%
\bibitem [{\citenamefont {Lin}\ \emph {et~al.}(2011)\citenamefont {Lin},
  \citenamefont {Jim{\'e}nez-Garc{\'\i}a},\ and\ \citenamefont
  {Spielman}}]{Lin:2011hn}%
  \BibitemOpen
  \bibfield  {author} {\bibinfo {author} {\bibfnamefont {Y.~J.}\ \bibnamefont
  {Lin}}, \bibinfo {author} {\bibfnamefont {K.}~\bibnamefont
  {Jim{\'e}nez-Garc{\'\i}a}}, \ and\ \bibinfo {author} {\bibfnamefont {I.~B.}\
  \bibnamefont {Spielman}},\ }\href@noop {} {\bibfield  {journal} {\bibinfo
  {journal} {Nature}\ }\textbf {\bibinfo {volume} {471}},\ \bibinfo {pages}
  {83} (\bibinfo {year} {2011})}\BibitemShut {NoStop}%
\bibitem [{\citenamefont {Higbie}\ and\ \citenamefont
  {Stamper-Kurn}(2002)}]{Higbie:2002gb}%
  \BibitemOpen
  \bibfield  {author} {\bibinfo {author} {\bibfnamefont {J.}~\bibnamefont
  {Higbie}}\ and\ \bibinfo {author} {\bibfnamefont {D.~M.}\ \bibnamefont
  {Stamper-Kurn}},\ }\href@noop {} {\bibfield  {journal} {\bibinfo  {journal}
  {Physical Review Letters}\ }\textbf {\bibinfo {volume} {88}},\ \bibinfo
  {pages} {090401} (\bibinfo {year} {2002})}\BibitemShut {NoStop}%
\bibitem [{\citenamefont {Liu}\ \emph {et~al.}(2009)\citenamefont {Liu},
  \citenamefont {Borunda}, \citenamefont {Liu},\ and\ \citenamefont
  {Sinova}}]{Liu:2009hj}%
  \BibitemOpen
  \bibfield  {author} {\bibinfo {author} {\bibfnamefont {X.-J.}\ \bibnamefont
  {Liu}}, \bibinfo {author} {\bibfnamefont {M.}~\bibnamefont {Borunda}},
  \bibinfo {author} {\bibfnamefont {X.}~\bibnamefont {Liu}}, \ and\ \bibinfo
  {author} {\bibfnamefont {J.}~\bibnamefont {Sinova}},\ }\href@noop {}
  {\bibfield  {journal} {\bibinfo  {journal} {Physical Review Letters}\
  }\textbf {\bibinfo {volume} {102}} (\bibinfo {year} {2009})}\BibitemShut
  {NoStop}%
\bibitem [{\citenamefont {Spielman}(2009)}]{Spielman:2009ej}%
  \BibitemOpen
  \bibfield  {author} {\bibinfo {author} {\bibfnamefont {I.~B.}\ \bibnamefont
  {Spielman}},\ }\href@noop {} {\bibfield  {journal} {\bibinfo  {journal}
  {Physical Review A}\ }\textbf {\bibinfo {volume} {79}},\ \bibinfo {pages}
  {063613} (\bibinfo {year} {2009})}\BibitemShut {NoStop}%
\bibitem [{\citenamefont {Luo}\ \emph {et~al.}(2016)\citenamefont {Luo},
  \citenamefont {Wu}, \citenamefont {Chen}, \citenamefont {Guan}, \citenamefont
  {Gao}, \citenamefont {Xu}, \citenamefont {You},\ and\ \citenamefont
  {Wang}}]{Luo:2016iw}%
  \BibitemOpen
  \bibfield  {author} {\bibinfo {author} {\bibfnamefont {X.}~\bibnamefont
  {Luo}}, \bibinfo {author} {\bibfnamefont {L.}~\bibnamefont {Wu}}, \bibinfo
  {author} {\bibfnamefont {J.}~\bibnamefont {Chen}}, \bibinfo {author}
  {\bibfnamefont {Q.}~\bibnamefont {Guan}}, \bibinfo {author} {\bibfnamefont
  {K.}~\bibnamefont {Gao}}, \bibinfo {author} {\bibfnamefont {Z.-F.}\
  \bibnamefont {Xu}}, \bibinfo {author} {\bibfnamefont {L.}~\bibnamefont
  {You}}, \ and\ \bibinfo {author} {\bibfnamefont {R.}~\bibnamefont {Wang}},\
  }\href@noop {} {\bibfield  {journal} {\bibinfo  {journal} {Scientific
  reports}\ }\textbf {\bibinfo {volume} {6}},\ \bibinfo {pages} {18983}
  (\bibinfo {year} {2016})}\BibitemShut {NoStop}%
\bibitem [{\citenamefont {Jim\'enez-Garc\'{\i}a}\ \emph
  {et~al.}(2015)\citenamefont {Jim\'enez-Garc\'{\i}a}, \citenamefont {LeBlanc},
  \citenamefont {Williams}, \citenamefont {Beeler}, \citenamefont {Qu},
  \citenamefont {Gong}, \citenamefont {Zhang},\ and\ \citenamefont
  {Spielman}}]{Jimenez-Garcia2015}%
  \BibitemOpen
  \bibfield  {author} {\bibinfo {author} {\bibfnamefont {K.}~\bibnamefont
  {Jim\'enez-Garc\'{\i}a}}, \bibinfo {author} {\bibfnamefont {L.~J.}\
  \bibnamefont {LeBlanc}}, \bibinfo {author} {\bibfnamefont {R.~A.}\
  \bibnamefont {Williams}}, \bibinfo {author} {\bibfnamefont {M.~C.}\
  \bibnamefont {Beeler}}, \bibinfo {author} {\bibfnamefont {C.}~\bibnamefont
  {Qu}}, \bibinfo {author} {\bibfnamefont {M.}~\bibnamefont {Gong}}, \bibinfo
  {author} {\bibfnamefont {C.}~\bibnamefont {Zhang}}, \ and\ \bibinfo {author}
  {\bibfnamefont {I.~B.}\ \bibnamefont {Spielman}},\ }\href {\doibase
  10.1103/PhysRevLett.114.125301} {\bibfield  {journal} {\bibinfo  {journal}
  {Phys. Rev. Lett.}\ }\textbf {\bibinfo {volume} {114}},\ \bibinfo {pages}
  {125301} (\bibinfo {year} {2015})}\BibitemShut {NoStop}%
\bibitem [{\citenamefont {Williams}\ \emph {et~al.}(2012)\citenamefont
  {Williams}, \citenamefont {LeBlanc}, \citenamefont {Jim{\'e}nez-Garc{\'\i}a},
  \citenamefont {Beeler}, \citenamefont {Perry}, \citenamefont {Phillips},\
  and\ \citenamefont {Spielman}}]{Williams:2012gs}%
  \BibitemOpen
  \bibfield  {author} {\bibinfo {author} {\bibfnamefont {R.~A.}\ \bibnamefont
  {Williams}}, \bibinfo {author} {\bibfnamefont {L.~J.}\ \bibnamefont
  {LeBlanc}}, \bibinfo {author} {\bibfnamefont {K.}~\bibnamefont
  {Jim{\'e}nez-Garc{\'\i}a}}, \bibinfo {author} {\bibfnamefont {M.~C.}\
  \bibnamefont {Beeler}}, \bibinfo {author} {\bibfnamefont {A.~R.}\
  \bibnamefont {Perry}}, \bibinfo {author} {\bibfnamefont {W.~D.}\ \bibnamefont
  {Phillips}}, \ and\ \bibinfo {author} {\bibfnamefont {I.~B.}\ \bibnamefont
  {Spielman}},\ }\href@noop {} {\bibfield  {journal} {\bibinfo  {journal}
  {Science}\ }\textbf {\bibinfo {volume} {335}},\ \bibinfo {pages} {314}
  (\bibinfo {year} {2012})}\BibitemShut {NoStop}%
\bibitem [{\citenamefont {Zhang}\ \emph {et~al.}(2012)\citenamefont {Zhang},
  \citenamefont {Ji}, \citenamefont {Chen}, \citenamefont {Zhang},
  \citenamefont {Du}, \citenamefont {Yan}, \citenamefont {Pan}, \citenamefont
  {Zhao}, \citenamefont {Deng}, \citenamefont {Zhai}, \citenamefont {Chen},\
  and\ \citenamefont {Pan}}]{Zhang2012}%
  \BibitemOpen
  \bibfield  {author} {\bibinfo {author} {\bibfnamefont {J.-Y.}\ \bibnamefont
  {Zhang}}, \bibinfo {author} {\bibfnamefont {S.-C.}\ \bibnamefont {Ji}},
  \bibinfo {author} {\bibfnamefont {Z.}~\bibnamefont {Chen}}, \bibinfo {author}
  {\bibfnamefont {L.}~\bibnamefont {Zhang}}, \bibinfo {author} {\bibfnamefont
  {Z.-D.}\ \bibnamefont {Du}}, \bibinfo {author} {\bibfnamefont
  {B.}~\bibnamefont {Yan}}, \bibinfo {author} {\bibfnamefont {G.-S.}\
  \bibnamefont {Pan}}, \bibinfo {author} {\bibfnamefont {B.}~\bibnamefont
  {Zhao}}, \bibinfo {author} {\bibfnamefont {Y.-J.}\ \bibnamefont {Deng}},
  \bibinfo {author} {\bibfnamefont {H.}~\bibnamefont {Zhai}}, \bibinfo {author}
  {\bibfnamefont {S.}~\bibnamefont {Chen}}, \ and\ \bibinfo {author}
  {\bibfnamefont {J.-W.}\ \bibnamefont {Pan}},\ }\href {\doibase
  10.1103/PhysRevLett.109.115301} {\bibfield  {journal} {\bibinfo  {journal}
  {Phys. Rev. Lett.}\ }\textbf {\bibinfo {volume} {109}},\ \bibinfo {pages}
  {115301} (\bibinfo {year} {2012})}\BibitemShut {NoStop}%
\bibitem [{\citenamefont {Qu}\ \emph {et~al.}(2013)\citenamefont {Qu},
  \citenamefont {Hamner}, \citenamefont {Gong}, \citenamefont {Zhang},\ and\
  \citenamefont {Engels}}]{Qu2013}%
  \BibitemOpen
  \bibfield  {author} {\bibinfo {author} {\bibfnamefont {C.}~\bibnamefont
  {Qu}}, \bibinfo {author} {\bibfnamefont {C.}~\bibnamefont {Hamner}}, \bibinfo
  {author} {\bibfnamefont {M.}~\bibnamefont {Gong}}, \bibinfo {author}
  {\bibfnamefont {C.}~\bibnamefont {Zhang}}, \ and\ \bibinfo {author}
  {\bibfnamefont {P.}~\bibnamefont {Engels}},\ }\href {\doibase
  10.1103/PhysRevA.88.021604} {\bibfield  {journal} {\bibinfo  {journal} {Phys.
  Rev. A}\ }\textbf {\bibinfo {volume} {88}},\ \bibinfo {pages} {021604}
  (\bibinfo {year} {2013})}\BibitemShut {NoStop}%
\bibitem [{\citenamefont {Hamner}\ \emph {et~al.}(2014)\citenamefont {Hamner},
  \citenamefont {Qu}, \citenamefont {Zhang}, \citenamefont {Chang},
  \citenamefont {Gong}, \citenamefont {Zhang},\ and\ \citenamefont
  {Engels}}]{Hamner2014}%
  \BibitemOpen
  \bibfield  {author} {\bibinfo {author} {\bibfnamefont {C.}~\bibnamefont
  {Hamner}}, \bibinfo {author} {\bibfnamefont {C.}~\bibnamefont {Qu}}, \bibinfo
  {author} {\bibfnamefont {Y.}~\bibnamefont {Zhang}}, \bibinfo {author}
  {\bibfnamefont {J.}~\bibnamefont {Chang}}, \bibinfo {author} {\bibfnamefont
  {M.}~\bibnamefont {Gong}}, \bibinfo {author} {\bibfnamefont {C.}~\bibnamefont
  {Zhang}}, \ and\ \bibinfo {author} {\bibfnamefont {P.}~\bibnamefont
  {Engels}},\ }\href {http://dx.doi.org/10.1038/ncomms5023} {\bibfield
  {journal} {\bibinfo  {journal} {Nat Commun}\ }\textbf {\bibinfo {volume}
  {5}},\  (\bibinfo {year} {2014})}\BibitemShut {NoStop}%
\bibitem [{\citenamefont {Ji}\ \emph {et~al.}(2014)\citenamefont {Ji},
  \citenamefont {Zhang}, \citenamefont {Zhang}, \citenamefont {Du},
  \citenamefont {Zheng}, \citenamefont {Deng}, \citenamefont {Zhai},
  \citenamefont {Chen},\ and\ \citenamefont {Pan}}]{Ji2014}%
  \BibitemOpen
  \bibfield  {author} {\bibinfo {author} {\bibfnamefont {S.-C.}\ \bibnamefont
  {Ji}}, \bibinfo {author} {\bibfnamefont {J.-Y.}\ \bibnamefont {Zhang}},
  \bibinfo {author} {\bibfnamefont {L.}~\bibnamefont {Zhang}}, \bibinfo
  {author} {\bibfnamefont {Z.-D.}\ \bibnamefont {Du}}, \bibinfo {author}
  {\bibfnamefont {W.}~\bibnamefont {Zheng}}, \bibinfo {author} {\bibfnamefont
  {Y.-J.}\ \bibnamefont {Deng}}, \bibinfo {author} {\bibfnamefont
  {H.}~\bibnamefont {Zhai}}, \bibinfo {author} {\bibfnamefont {S.}~\bibnamefont
  {Chen}}, \ and\ \bibinfo {author} {\bibfnamefont {J.-W.}\ \bibnamefont
  {Pan}},\ }\href {http://dx.doi.org/10.1038/nphys2905} {\bibfield  {journal}
  {\bibinfo  {journal} {Nat Phys}\ }\textbf {\bibinfo {volume} {10}},\ \bibinfo
  {pages} {314} (\bibinfo {year} {2014})}\BibitemShut {NoStop}%
\bibitem [{\citenamefont {Olson}\ \emph {et~al.}(2014)\citenamefont {Olson},
  \citenamefont {Wang}, \citenamefont {Niffenegger}, \citenamefont {Li},
  \citenamefont {Greene},\ and\ \citenamefont {Chen}}]{Olson2014}%
  \BibitemOpen
  \bibfield  {author} {\bibinfo {author} {\bibfnamefont {A.~J.}\ \bibnamefont
  {Olson}}, \bibinfo {author} {\bibfnamefont {S.-J.}\ \bibnamefont {Wang}},
  \bibinfo {author} {\bibfnamefont {R.~J.}\ \bibnamefont {Niffenegger}},
  \bibinfo {author} {\bibfnamefont {C.-H.}\ \bibnamefont {Li}}, \bibinfo
  {author} {\bibfnamefont {C.~H.}\ \bibnamefont {Greene}}, \ and\ \bibinfo
  {author} {\bibfnamefont {Y.~P.}\ \bibnamefont {Chen}},\ }\href {\doibase
  10.1103/PhysRevA.90.013616} {\bibfield  {journal} {\bibinfo  {journal} {Phys.
  Rev. A}\ }\textbf {\bibinfo {volume} {90}},\ \bibinfo {pages} {013616}
  (\bibinfo {year} {2014})}\BibitemShut {NoStop}%
\bibitem [{\citenamefont {Ji}\ \emph {et~al.}(2015)\citenamefont {Ji},
  \citenamefont {Zhang}, \citenamefont {Xu}, \citenamefont {Wu}, \citenamefont
  {Deng}, \citenamefont {Chen},\ and\ \citenamefont {Pan}}]{Ji2015}%
  \BibitemOpen
  \bibfield  {author} {\bibinfo {author} {\bibfnamefont {S.-C.}\ \bibnamefont
  {Ji}}, \bibinfo {author} {\bibfnamefont {L.}~\bibnamefont {Zhang}}, \bibinfo
  {author} {\bibfnamefont {X.-T.}\ \bibnamefont {Xu}}, \bibinfo {author}
  {\bibfnamefont {Z.}~\bibnamefont {Wu}}, \bibinfo {author} {\bibfnamefont
  {Y.}~\bibnamefont {Deng}}, \bibinfo {author} {\bibfnamefont {S.}~\bibnamefont
  {Chen}}, \ and\ \bibinfo {author} {\bibfnamefont {J.-W.}\ \bibnamefont
  {Pan}},\ }\href {\doibase 10.1103/PhysRevLett.114.105301} {\bibfield
  {journal} {\bibinfo  {journal} {Phys. Rev. Lett.}\ }\textbf {\bibinfo
  {volume} {114}},\ \bibinfo {pages} {105301} (\bibinfo {year}
  {2015})}\BibitemShut {NoStop}%
\bibitem [{\citenamefont {Wang}\ \emph {et~al.}(2012)\citenamefont {Wang},
  \citenamefont {Yu}, \citenamefont {Fu}, \citenamefont {Miao}, \citenamefont
  {Huang}, \citenamefont {Chai}, \citenamefont {Zhai},\ and\ \citenamefont
  {Zhang}}]{Wang2012}%
  \BibitemOpen
  \bibfield  {author} {\bibinfo {author} {\bibfnamefont {P.}~\bibnamefont
  {Wang}}, \bibinfo {author} {\bibfnamefont {Z.-Q.}\ \bibnamefont {Yu}},
  \bibinfo {author} {\bibfnamefont {Z.}~\bibnamefont {Fu}}, \bibinfo {author}
  {\bibfnamefont {J.}~\bibnamefont {Miao}}, \bibinfo {author} {\bibfnamefont
  {L.}~\bibnamefont {Huang}}, \bibinfo {author} {\bibfnamefont
  {S.}~\bibnamefont {Chai}}, \bibinfo {author} {\bibfnamefont {H.}~\bibnamefont
  {Zhai}}, \ and\ \bibinfo {author} {\bibfnamefont {J.}~\bibnamefont {Zhang}},\
  }\href {\doibase 10.1103/PhysRevLett.109.095301} {\bibfield  {journal}
  {\bibinfo  {journal} {Phys. Rev. Lett.}\ }\textbf {\bibinfo {volume} {109}},\
  \bibinfo {pages} {095301} (\bibinfo {year} {2012})}\BibitemShut {NoStop}%
\bibitem [{\citenamefont {Cheuk}\ \emph {et~al.}(2012)\citenamefont {Cheuk},
  \citenamefont {Sommer}, \citenamefont {Hadzibabic}, \citenamefont {Yefsah},
  \citenamefont {Bakr},\ and\ \citenamefont {Zwierlein}}]{2012PhRvL.109i5302C}%
  \BibitemOpen
  \bibfield  {author} {\bibinfo {author} {\bibfnamefont {L.~W.}\ \bibnamefont
  {Cheuk}}, \bibinfo {author} {\bibfnamefont {A.~T.}\ \bibnamefont {Sommer}},
  \bibinfo {author} {\bibfnamefont {Z.}~\bibnamefont {Hadzibabic}}, \bibinfo
  {author} {\bibfnamefont {T.}~\bibnamefont {Yefsah}}, \bibinfo {author}
  {\bibfnamefont {W.~S.}\ \bibnamefont {Bakr}}, \ and\ \bibinfo {author}
  {\bibfnamefont {M.~W.}\ \bibnamefont {Zwierlein}},\ }\href@noop {} {\bibfield
   {journal} {\bibinfo  {journal} {Physical Review Letters}\ }\textbf {\bibinfo
  {volume} {109}},\ \bibinfo {pages} {095302} (\bibinfo {year}
  {2012})}\BibitemShut {NoStop}%
\bibitem [{\citenamefont {Williams}\ \emph {et~al.}(2013)\citenamefont
  {Williams}, \citenamefont {Beeler}, \citenamefont {LeBlanc}, \citenamefont
  {Jim\'enez-Garc\'{\i}a},\ and\ \citenamefont {Spielman}}]{Williams2013}%
  \BibitemOpen
  \bibfield  {author} {\bibinfo {author} {\bibfnamefont {R.~A.}\ \bibnamefont
  {Williams}}, \bibinfo {author} {\bibfnamefont {M.~C.}\ \bibnamefont
  {Beeler}}, \bibinfo {author} {\bibfnamefont {L.~J.}\ \bibnamefont {LeBlanc}},
  \bibinfo {author} {\bibfnamefont {K.}~\bibnamefont {Jim\'enez-Garc\'{\i}a}},
  \ and\ \bibinfo {author} {\bibfnamefont {I.~B.}\ \bibnamefont {Spielman}},\
  }\href {\doibase 10.1103/PhysRevLett.111.095301} {\bibfield  {journal}
  {\bibinfo  {journal} {Phys. Rev. Lett.}\ }\textbf {\bibinfo {volume} {111}},\
  \bibinfo {pages} {095301} (\bibinfo {year} {2013})}\BibitemShut {NoStop}%
\bibitem [{\citenamefont {Fu}\ \emph {et~al.}(2014)\citenamefont {Fu},
  \citenamefont {Huang}, \citenamefont {Meng}, \citenamefont {Wang},
  \citenamefont {Zhang}, \citenamefont {Zhang}, \citenamefont {Zhai},
  \citenamefont {Zhang},\ and\ \citenamefont {Zhang}}]{Fu2014}%
  \BibitemOpen
  \bibfield  {author} {\bibinfo {author} {\bibfnamefont {Z.}~\bibnamefont
  {Fu}}, \bibinfo {author} {\bibfnamefont {L.}~\bibnamefont {Huang}}, \bibinfo
  {author} {\bibfnamefont {Z.}~\bibnamefont {Meng}}, \bibinfo {author}
  {\bibfnamefont {P.}~\bibnamefont {Wang}}, \bibinfo {author} {\bibfnamefont
  {L.}~\bibnamefont {Zhang}}, \bibinfo {author} {\bibfnamefont
  {S.}~\bibnamefont {Zhang}}, \bibinfo {author} {\bibfnamefont
  {H.}~\bibnamefont {Zhai}}, \bibinfo {author} {\bibfnamefont {P.}~\bibnamefont
  {Zhang}}, \ and\ \bibinfo {author} {\bibfnamefont {J.}~\bibnamefont
  {Zhang}},\ }\href {http://dx.doi.org/10.1038/nphys2824} {\bibfield  {journal}
  {\bibinfo  {journal} {Nat Phys}\ }\textbf {\bibinfo {volume} {10}},\ \bibinfo
  {pages} {110} (\bibinfo {year} {2014})}\BibitemShut {NoStop}%
\bibitem [{\citenamefont {Nathaniel Q.~Burdick}\ and\ \citenamefont
  {Lev}(2016)}]{NathanieQ2016}%
  \BibitemOpen
  \bibfield  {author} {\bibinfo {author} {\bibfnamefont {Y.~T.}\ \bibnamefont
  {Nathaniel Q.~Burdick}}\ and\ \bibinfo {author} {\bibfnamefont {B.~L.}\
  \bibnamefont {Lev}},\ }\href@noop {} {\bibfield  {journal} {\bibinfo
  {journal} {arXiv:1605.03211}\ } (\bibinfo {year} {2016})}\BibitemShut
  {NoStop}%
\bibitem [{\citenamefont {Wu}\ \emph {et~al.}(2016)\citenamefont {Wu},
  \citenamefont {Zhang}, \citenamefont {Sun}, \citenamefont {Xu}, \citenamefont
  {Wang}, \citenamefont {Ji}, \citenamefont {Deng}, \citenamefont {Chen},
  \citenamefont {Liu},\ and\ \citenamefont {Pan}}]{2015arXiv151108170W}%
  \BibitemOpen
  \bibfield  {author} {\bibinfo {author} {\bibfnamefont {Z.}~\bibnamefont
  {Wu}}, \bibinfo {author} {\bibfnamefont {L.}~\bibnamefont {Zhang}}, \bibinfo
  {author} {\bibfnamefont {W.}~\bibnamefont {Sun}}, \bibinfo {author}
  {\bibfnamefont {X.-T.}\ \bibnamefont {Xu}}, \bibinfo {author} {\bibfnamefont
  {B.-Z.}\ \bibnamefont {Wang}}, \bibinfo {author} {\bibfnamefont {S.-C.}\
  \bibnamefont {Ji}}, \bibinfo {author} {\bibfnamefont {Y.}~\bibnamefont
  {Deng}}, \bibinfo {author} {\bibfnamefont {S.}~\bibnamefont {Chen}}, \bibinfo
  {author} {\bibfnamefont {X.-J.}\ \bibnamefont {Liu}}, \ and\ \bibinfo
  {author} {\bibfnamefont {J.-W.}\ \bibnamefont {Pan}},\ }\href@noop {}
  {\bibfield  {journal} {\bibinfo  {journal} {Science}\ }\textbf {\bibinfo
  {volume} {354}},\ \bibinfo {pages} {83} (\bibinfo {year} {2016})}\BibitemShut
  {NoStop}%
\bibitem [{\citenamefont {Huang}\ \emph {et~al.}(2016)\citenamefont {Huang},
  \citenamefont {Meng}, \citenamefont {Wang}, \citenamefont {Peng},
  \citenamefont {Zhang}, \citenamefont {Chen}, \citenamefont {Li},
  \citenamefont {Zhou},\ and\ \citenamefont {Zhang}}]{Huang2016}%
  \BibitemOpen
  \bibfield  {author} {\bibinfo {author} {\bibfnamefont {L.}~\bibnamefont
  {Huang}}, \bibinfo {author} {\bibfnamefont {Z.}~\bibnamefont {Meng}},
  \bibinfo {author} {\bibfnamefont {P.}~\bibnamefont {Wang}}, \bibinfo {author}
  {\bibfnamefont {P.}~\bibnamefont {Peng}}, \bibinfo {author} {\bibfnamefont
  {S.-L.}\ \bibnamefont {Zhang}}, \bibinfo {author} {\bibfnamefont
  {L.}~\bibnamefont {Chen}}, \bibinfo {author} {\bibfnamefont {D.}~\bibnamefont
  {Li}}, \bibinfo {author} {\bibfnamefont {Q.}~\bibnamefont {Zhou}}, \ and\
  \bibinfo {author} {\bibfnamefont {J.}~\bibnamefont {Zhang}},\ }\href
  {http://dx.doi.org/10.1038/nphys3672} {\bibfield  {journal} {\bibinfo
  {journal} {Nat Phys}\ }\textbf {\bibinfo {volume} {12}},\ \bibinfo {pages}
  {540} (\bibinfo {year} {2016})}\BibitemShut {NoStop}%
\bibitem [{\citenamefont {Cui}\ \emph {et~al.}(2013{\natexlab{a}})\citenamefont
  {Cui}, \citenamefont {Lian}, \citenamefont {Ho}, \citenamefont {Lev},\ and\
  \citenamefont {Zhai}}]{Cui2013}%
  \BibitemOpen
  \bibfield  {author} {\bibinfo {author} {\bibfnamefont {X.}~\bibnamefont
  {Cui}}, \bibinfo {author} {\bibfnamefont {B.}~\bibnamefont {Lian}}, \bibinfo
  {author} {\bibfnamefont {T.-L.}\ \bibnamefont {Ho}}, \bibinfo {author}
  {\bibfnamefont {B.~L.}\ \bibnamefont {Lev}}, \ and\ \bibinfo {author}
  {\bibfnamefont {H.}~\bibnamefont {Zhai}},\ }\href {\doibase
  10.1103/PhysRevA.88.011601} {\bibfield  {journal} {\bibinfo  {journal} {Phys.
  Rev. A}\ }\textbf {\bibinfo {volume} {88}},\ \bibinfo {pages} {011601}
  (\bibinfo {year} {2013}{\natexlab{a}})}\BibitemShut {NoStop}%
\bibitem [{\citenamefont {Wall}\ \emph {et~al.}(2016)\citenamefont {Wall},
  \citenamefont {Koller}, \citenamefont {Li}, \citenamefont {Zhang},
  \citenamefont {Cooper}, \citenamefont {Ye},\ and\ \citenamefont
  {Rey}}]{Wall2016}%
  \BibitemOpen
  \bibfield  {author} {\bibinfo {author} {\bibfnamefont {M.~L.}\ \bibnamefont
  {Wall}}, \bibinfo {author} {\bibfnamefont {A.~P.}\ \bibnamefont {Koller}},
  \bibinfo {author} {\bibfnamefont {S.}~\bibnamefont {Li}}, \bibinfo {author}
  {\bibfnamefont {X.}~\bibnamefont {Zhang}}, \bibinfo {author} {\bibfnamefont
  {N.~R.}\ \bibnamefont {Cooper}}, \bibinfo {author} {\bibfnamefont
  {J.}~\bibnamefont {Ye}}, \ and\ \bibinfo {author} {\bibfnamefont {A.~M.}\
  \bibnamefont {Rey}},\ }\href {\doibase 10.1103/PhysRevLett.116.035301}
  {\bibfield  {journal} {\bibinfo  {journal} {Phys. Rev. Lett.}\ }\textbf
  {\bibinfo {volume} {116}},\ \bibinfo {pages} {035301} (\bibinfo {year}
  {2016})}\BibitemShut {NoStop}%
\bibitem [{\citenamefont {Atala}\ \emph {et~al.}(2014)\citenamefont {Atala},
  \citenamefont {Aidelsburger}, \citenamefont {Lohse}, \citenamefont
  {Barreiro}, \citenamefont {Paredes},\ and\ \citenamefont
  {Bloch}}]{Atala:2014gf}%
  \BibitemOpen
  \bibfield  {author} {\bibinfo {author} {\bibfnamefont {M.}~\bibnamefont
  {Atala}}, \bibinfo {author} {\bibfnamefont {M.}~\bibnamefont {Aidelsburger}},
  \bibinfo {author} {\bibfnamefont {M.}~\bibnamefont {Lohse}}, \bibinfo
  {author} {\bibfnamefont {J.~T.}\ \bibnamefont {Barreiro}}, \bibinfo {author}
  {\bibfnamefont {B.}~\bibnamefont {Paredes}}, \ and\ \bibinfo {author}
  {\bibfnamefont {I.}~\bibnamefont {Bloch}},\ }\href@noop {} {\bibfield
  {journal} {\bibinfo  {journal} {Nat.Phys}\ }\textbf {\bibinfo {volume}
  {10}},\ \bibinfo {pages} {588} (\bibinfo {year} {2014})}\BibitemShut
  {NoStop}%
\bibitem [{\citenamefont {Li}\ \emph {et~al.}(2016)\citenamefont {Li},
  \citenamefont {Huang}, \citenamefont {Shteynas}, \citenamefont {Burchesky},
  \citenamefont {Top}, \citenamefont {Su}, \citenamefont {Lee}, \citenamefont
  {Jamison},\ and\ \citenamefont {Ketterle}}]{Li:2016vc}%
  \BibitemOpen
  \bibfield  {author} {\bibinfo {author} {\bibfnamefont {J.}~\bibnamefont
  {Li}}, \bibinfo {author} {\bibfnamefont {W.}~\bibnamefont {Huang}}, \bibinfo
  {author} {\bibfnamefont {B.}~\bibnamefont {Shteynas}}, \bibinfo {author}
  {\bibfnamefont {S.}~\bibnamefont {Burchesky}}, \bibinfo {author}
  {\bibfnamefont {F.~C.}\ \bibnamefont {Top}}, \bibinfo {author} {\bibfnamefont
  {E.}~\bibnamefont {Su}}, \bibinfo {author} {\bibfnamefont {J.}~\bibnamefont
  {Lee}}, \bibinfo {author} {\bibfnamefont {A.~O.}\ \bibnamefont {Jamison}}, \
  and\ \bibinfo {author} {\bibfnamefont {W.}~\bibnamefont {Ketterle}},\
  }\href@noop {} {\  (\bibinfo {year} {2016})},\ \Eprint
  {http://arxiv.org/abs/1606.03514} {1606.03514} \BibitemShut {NoStop}%
\bibitem [{\citenamefont {Daley}(2011)}]{2011arXiv1106.5712D}%
  \BibitemOpen
  \bibfield  {author} {\bibinfo {author} {\bibfnamefont {A.~J.}\ \bibnamefont
  {Daley}},\ }\href@noop {} {\bibfield  {journal} {\bibinfo  {journal} {Quantum
  Information Processing}\ }\textbf {\bibinfo {volume} {10}},\ \bibinfo {pages}
  {865} (\bibinfo {year} {2011})}\BibitemShut {NoStop}%
\bibitem [{\citenamefont {Cazalilla}\ and\ \citenamefont
  {Rey}(2014)}]{Cazalilla:2014ut}%
  \BibitemOpen
  \bibfield  {author} {\bibinfo {author} {\bibfnamefont {M.~A.}\ \bibnamefont
  {Cazalilla}}\ and\ \bibinfo {author} {\bibfnamefont {A.~M.}\ \bibnamefont
  {Rey}},\ }\href@noop {} {\bibfield  {journal} {\bibinfo  {journal} {Reports
  on Progress in Physics}\ }\textbf {\bibinfo {volume} {77}},\ \bibinfo {pages}
  {124401} (\bibinfo {year} {2014})}\BibitemShut {NoStop}%
\bibitem [{\citenamefont {Pagano}\ \emph {et~al.}(2014)\citenamefont {Pagano},
  \citenamefont {Mancini}, \citenamefont {Cappellini}, \citenamefont
  {Lombardi}, \citenamefont {Sch{\"a}fer}, \citenamefont {Hu}, \citenamefont
  {Liu}, \citenamefont {Catani}, \citenamefont {Sias}, \citenamefont
  {Inguscio},\ and\ \citenamefont {Fallani}}]{Pagano:2014hy}%
  \BibitemOpen
  \bibfield  {author} {\bibinfo {author} {\bibfnamefont {G.}~\bibnamefont
  {Pagano}}, \bibinfo {author} {\bibfnamefont {M.}~\bibnamefont {Mancini}},
  \bibinfo {author} {\bibfnamefont {G.}~\bibnamefont {Cappellini}}, \bibinfo
  {author} {\bibfnamefont {P.}~\bibnamefont {Lombardi}}, \bibinfo {author}
  {\bibfnamefont {F.}~\bibnamefont {Sch{\"a}fer}}, \bibinfo {author}
  {\bibfnamefont {H.}~\bibnamefont {Hu}}, \bibinfo {author} {\bibfnamefont
  {X.-J.}\ \bibnamefont {Liu}}, \bibinfo {author} {\bibfnamefont
  {J.}~\bibnamefont {Catani}}, \bibinfo {author} {\bibfnamefont
  {C.}~\bibnamefont {Sias}}, \bibinfo {author} {\bibfnamefont {M.}~\bibnamefont
  {Inguscio}}, \ and\ \bibinfo {author} {\bibfnamefont {L.}~\bibnamefont
  {Fallani}},\ }\href@noop {} {\bibfield  {journal} {\bibinfo  {journal}
  {Nature Physics}\ }\textbf {\bibinfo {volume} {10}},\ \bibinfo {pages} {198}
  (\bibinfo {year} {2014})}\BibitemShut {NoStop}%
\bibitem [{\citenamefont {Taie}\ \emph {et~al.}(2012)\citenamefont {Taie},
  \citenamefont {Yamazaki}, \citenamefont {Sugawa},\ and\ \citenamefont
  {Takahashi}}]{Taie:2012vp}%
  \BibitemOpen
  \bibfield  {author} {\bibinfo {author} {\bibfnamefont {S.}~\bibnamefont
  {Taie}}, \bibinfo {author} {\bibfnamefont {R.}~\bibnamefont {Yamazaki}},
  \bibinfo {author} {\bibfnamefont {S.}~\bibnamefont {Sugawa}}, \ and\ \bibinfo
  {author} {\bibfnamefont {Y.}~\bibnamefont {Takahashi}},\ }\href@noop {}
  {\bibfield  {journal} {\bibinfo  {journal} {Nat.Phys}\ }\textbf {\bibinfo
  {volume} {8}},\ \bibinfo {pages} {825} (\bibinfo {year} {2012})}\BibitemShut
  {NoStop}%
\bibitem [{\citenamefont {Mancini}\ \emph
  {et~al.}(2015{\natexlab{a}})\citenamefont {Mancini}, \citenamefont {Pagano},
  \citenamefont {Cappellini}, \citenamefont {Livi}, \citenamefont {Rider},
  \citenamefont {Catani}, \citenamefont {Sias}, \citenamefont {Zoller},
  \citenamefont {Inguscio}, \citenamefont {Dalmonte},\ and\ \citenamefont
  {Fallani}}]{Mancini2015}%
  \BibitemOpen
  \bibfield  {author} {\bibinfo {author} {\bibfnamefont {M.}~\bibnamefont
  {Mancini}}, \bibinfo {author} {\bibfnamefont {G.}~\bibnamefont {Pagano}},
  \bibinfo {author} {\bibfnamefont {G.}~\bibnamefont {Cappellini}}, \bibinfo
  {author} {\bibfnamefont {L.}~\bibnamefont {Livi}}, \bibinfo {author}
  {\bibfnamefont {M.}~\bibnamefont {Rider}}, \bibinfo {author} {\bibfnamefont
  {J.}~\bibnamefont {Catani}}, \bibinfo {author} {\bibfnamefont
  {C.}~\bibnamefont {Sias}}, \bibinfo {author} {\bibfnamefont {P.}~\bibnamefont
  {Zoller}}, \bibinfo {author} {\bibfnamefont {M.}~\bibnamefont {Inguscio}},
  \bibinfo {author} {\bibfnamefont {M.}~\bibnamefont {Dalmonte}}, \ and\
  \bibinfo {author} {\bibfnamefont {L.}~\bibnamefont {Fallani}},\ }\href
  {\doibase 10.1126/science.aaa8736} {\bibfield  {journal} {\bibinfo  {journal}
  {Science}\ }\textbf {\bibinfo {volume} {349}},\ \bibinfo {pages} {1510}
  (\bibinfo {year} {2015}{\natexlab{a}})}\BibitemShut {NoStop}%
\bibitem [{\citenamefont {Nakajima}\ \emph {et~al.}(2016)\citenamefont
  {Nakajima}, \citenamefont {Tomita}, \citenamefont {Taie}, \citenamefont
  {Ichinose}, \citenamefont {Ozawa}, \citenamefont {Wang}, \citenamefont
  {Troyer},\ and\ \citenamefont {Takahashi}}]{2016NatPh..12..296N}%
  \BibitemOpen
  \bibfield  {author} {\bibinfo {author} {\bibfnamefont {S.}~\bibnamefont
  {Nakajima}}, \bibinfo {author} {\bibfnamefont {T.}~\bibnamefont {Tomita}},
  \bibinfo {author} {\bibfnamefont {S.}~\bibnamefont {Taie}}, \bibinfo {author}
  {\bibfnamefont {T.}~\bibnamefont {Ichinose}}, \bibinfo {author}
  {\bibfnamefont {H.}~\bibnamefont {Ozawa}}, \bibinfo {author} {\bibfnamefont
  {L.}~\bibnamefont {Wang}}, \bibinfo {author} {\bibfnamefont {M.}~\bibnamefont
  {Troyer}}, \ and\ \bibinfo {author} {\bibfnamefont {Y.}~\bibnamefont
  {Takahashi}},\ }\href@noop {} {\bibfield  {journal} {\bibinfo  {journal}
  {Nat.Phys}\ }\textbf {\bibinfo {volume} {12}},\ \bibinfo {pages} {296}
  (\bibinfo {year} {2016})}\BibitemShut {NoStop}%
\bibitem [{\citenamefont {Wu}\ \emph {et~al.}(2003)\citenamefont {Wu},
  \citenamefont {Hu},\ and\ \citenamefont {Zhang}}]{Wu:2003es}%
  \BibitemOpen
  \bibfield  {author} {\bibinfo {author} {\bibfnamefont {C.}~\bibnamefont
  {Wu}}, \bibinfo {author} {\bibfnamefont {J.-p.}\ \bibnamefont {Hu}}, \ and\
  \bibinfo {author} {\bibfnamefont {S.-C.}\ \bibnamefont {Zhang}},\ }\href@noop
  {} {\bibfield  {journal} {\bibinfo  {journal} {Physical Review Lettres}\
  }\textbf {\bibinfo {volume} {91}},\ \bibinfo {pages} {186402} (\bibinfo
  {year} {2003})}\BibitemShut {NoStop}%
\bibitem [{\citenamefont {Gorshkov}\ \emph {et~al.}(2010)\citenamefont
  {Gorshkov}, \citenamefont {Hermele}, \citenamefont {Gurarie}, \citenamefont
  {Xu}, \citenamefont {Julienne}, \citenamefont {Ye}, \citenamefont {Zoller},
  \citenamefont {Demler}, \citenamefont {Lukin},\ and\ \citenamefont
  {Rey}}]{Gorshkov:2010hw}%
  \BibitemOpen
  \bibfield  {author} {\bibinfo {author} {\bibfnamefont {A.~V.}\ \bibnamefont
  {Gorshkov}}, \bibinfo {author} {\bibfnamefont {M.}~\bibnamefont {Hermele}},
  \bibinfo {author} {\bibfnamefont {V.}~\bibnamefont {Gurarie}}, \bibinfo
  {author} {\bibfnamefont {C.}~\bibnamefont {Xu}}, \bibinfo {author}
  {\bibfnamefont {P.~S.}\ \bibnamefont {Julienne}}, \bibinfo {author}
  {\bibfnamefont {J.}~\bibnamefont {Ye}}, \bibinfo {author} {\bibfnamefont
  {P.}~\bibnamefont {Zoller}}, \bibinfo {author} {\bibfnamefont
  {E.}~\bibnamefont {Demler}}, \bibinfo {author} {\bibfnamefont {M.~D.}\
  \bibnamefont {Lukin}}, \ and\ \bibinfo {author} {\bibfnamefont {A.~M.}\
  \bibnamefont {Rey}},\ }\href@noop {} {\bibfield  {journal} {\bibinfo
  {journal} {Nature Physics}\ }\textbf {\bibinfo {volume} {6}},\ \bibinfo
  {pages} {289} (\bibinfo {year} {2010})}\BibitemShut {NoStop}%
\bibitem [{\citenamefont {Cai}\ \emph {et~al.}(2013)\citenamefont {Cai},
  \citenamefont {Hung}, \citenamefont {Wang}, \citenamefont {Zheng},\ and\
  \citenamefont {Wu}}]{Cai:2013ke}%
  \BibitemOpen
  \bibfield  {author} {\bibinfo {author} {\bibfnamefont {Z.}~\bibnamefont
  {Cai}}, \bibinfo {author} {\bibfnamefont {H.-H.}\ \bibnamefont {Hung}},
  \bibinfo {author} {\bibfnamefont {L.}~\bibnamefont {Wang}}, \bibinfo {author}
  {\bibfnamefont {D.}~\bibnamefont {Zheng}}, \ and\ \bibinfo {author}
  {\bibfnamefont {C.}~\bibnamefont {Wu}},\ }\href@noop {} {\bibfield  {journal}
  {\bibinfo  {journal} {Physical Review Letters}\ }\textbf {\bibinfo {volume}
  {110}},\ \bibinfo {pages} {220401} (\bibinfo {year} {2013})}\BibitemShut
  {NoStop}%
\bibitem [{\citenamefont {Barbarino}\ \emph {et~al.}(2015)\citenamefont
  {Barbarino}, \citenamefont {Taddia}, \citenamefont {Rossini}, \citenamefont
  {Mazza},\ and\ \citenamefont {Fazio}}]{Barbarino:2015fi}%
  \BibitemOpen
  \bibfield  {author} {\bibinfo {author} {\bibfnamefont {S.}~\bibnamefont
  {Barbarino}}, \bibinfo {author} {\bibfnamefont {L.}~\bibnamefont {Taddia}},
  \bibinfo {author} {\bibfnamefont {D.}~\bibnamefont {Rossini}}, \bibinfo
  {author} {\bibfnamefont {L.}~\bibnamefont {Mazza}}, \ and\ \bibinfo {author}
  {\bibfnamefont {R.}~\bibnamefont {Fazio}},\ }\href@noop {} {\bibfield
  {journal} {\bibinfo  {journal} {Nature Communications}\ }\textbf {\bibinfo
  {volume} {6}},\ \bibinfo {pages} {8134} (\bibinfo {year} {2015})}\BibitemShut
  {NoStop}%
\bibitem [{\citenamefont {Banerjee}\ \emph {et~al.}(2013)\citenamefont
  {Banerjee}, \citenamefont {B{\"o}gli}, \citenamefont {Dalmonte},
  \citenamefont {Rico}, \citenamefont {Stebler}, \citenamefont {Wiese},\ and\
  \citenamefont {Zoller}}]{Banerjee:2013fl}%
  \BibitemOpen
  \bibfield  {author} {\bibinfo {author} {\bibfnamefont {D.}~\bibnamefont
  {Banerjee}}, \bibinfo {author} {\bibfnamefont {M.}~\bibnamefont {B{\"o}gli}},
  \bibinfo {author} {\bibfnamefont {M.}~\bibnamefont {Dalmonte}}, \bibinfo
  {author} {\bibfnamefont {E.}~\bibnamefont {Rico}}, \bibinfo {author}
  {\bibfnamefont {P.}~\bibnamefont {Stebler}}, \bibinfo {author} {\bibfnamefont
  {U.~J.}\ \bibnamefont {Wiese}}, \ and\ \bibinfo {author} {\bibfnamefont
  {P.}~\bibnamefont {Zoller}},\ }\href@noop {} {\bibfield  {journal} {\bibinfo
  {journal} {Physical Review Letters}\ }\textbf {\bibinfo {volume} {110}},\
  \bibinfo {pages} {125303} (\bibinfo {year} {2013})}\BibitemShut {NoStop}%
\bibitem [{\citenamefont {Fukuhara}\ \emph {et~al.}(2007)\citenamefont
  {Fukuhara}, \citenamefont {Takasu}, \citenamefont {Kumakura},\ and\
  \citenamefont {Takahashi}}]{Fukuhara:2007wg}%
  \BibitemOpen
  \bibfield  {author} {\bibinfo {author} {\bibfnamefont {T.}~\bibnamefont
  {Fukuhara}}, \bibinfo {author} {\bibfnamefont {Y.}~\bibnamefont {Takasu}},
  \bibinfo {author} {\bibfnamefont {M.}~\bibnamefont {Kumakura}}, \ and\
  \bibinfo {author} {\bibfnamefont {Y.}~\bibnamefont {Takahashi}},\ }\href@noop
  {} {\bibfield  {journal} {\bibinfo  {journal} {Physical Review Letters}\
  }\textbf {\bibinfo {volume} {98}},\ \bibinfo {pages} {030401} (\bibinfo
  {year} {2007})}\BibitemShut {NoStop}%
\bibitem [{\citenamefont {Stellmer}\ \emph {et~al.}(2011)\citenamefont
  {Stellmer}, \citenamefont {Grimm},\ and\ \citenamefont
  {SCHRECK}}]{Stellmer:2011iv}%
  \BibitemOpen
  \bibfield  {author} {\bibinfo {author} {\bibfnamefont {S.}~\bibnamefont
  {Stellmer}}, \bibinfo {author} {\bibfnamefont {R.}~\bibnamefont {Grimm}}, \
  and\ \bibinfo {author} {\bibfnamefont {F.}~\bibnamefont {SCHRECK}},\
  }\href@noop {} {\bibfield  {journal} {\bibinfo  {journal} {Physical Review
  A}\ }\textbf {\bibinfo {volume} {84}},\ \bibinfo {pages} {043611} (\bibinfo
  {year} {2011})}\BibitemShut {NoStop}%
\bibitem [{\citenamefont {Mancini}\ \emph
  {et~al.}(2015{\natexlab{b}})\citenamefont {Mancini}, \citenamefont {Pagano},
  \citenamefont {Cappellini}, \citenamefont {Livi}, \citenamefont {Rider},
  \citenamefont {Catani}, \citenamefont {Sias}, \citenamefont {Zoller},
  \citenamefont {Inguscio}, \citenamefont {Dalmonte},\ and\ \citenamefont
  {Fallani}}]{2015arXiv150202495M}%
  \BibitemOpen
  \bibfield  {author} {\bibinfo {author} {\bibfnamefont {M.}~\bibnamefont
  {Mancini}}, \bibinfo {author} {\bibfnamefont {G.}~\bibnamefont {Pagano}},
  \bibinfo {author} {\bibfnamefont {G.}~\bibnamefont {Cappellini}}, \bibinfo
  {author} {\bibfnamefont {L.}~\bibnamefont {Livi}}, \bibinfo {author}
  {\bibfnamefont {M.}~\bibnamefont {Rider}}, \bibinfo {author} {\bibfnamefont
  {J.}~\bibnamefont {Catani}}, \bibinfo {author} {\bibfnamefont
  {C.}~\bibnamefont {Sias}}, \bibinfo {author} {\bibfnamefont {P.}~\bibnamefont
  {Zoller}}, \bibinfo {author} {\bibfnamefont {M.}~\bibnamefont {Inguscio}},
  \bibinfo {author} {\bibfnamefont {M.}~\bibnamefont {Dalmonte}}, \ and\
  \bibinfo {author} {\bibfnamefont {L.}~\bibnamefont {Fallani}},\ }\href@noop
  {} {\bibfield  {journal} {\bibinfo  {journal} {Science}\ }\textbf {\bibinfo
  {volume} {349}},\ \bibinfo {pages} {1510} (\bibinfo {year}
  {2015}{\natexlab{b}})}\BibitemShut {NoStop}%
\bibitem [{\citenamefont {Taie}\ \emph {et~al.}(2010)\citenamefont {Taie},
  \citenamefont {Takasu}, \citenamefont {Sugawa}, \citenamefont {Yamazaki},
  \citenamefont {Tsujimoto}, \citenamefont {Murakami},\ and\ \citenamefont
  {Takahashi}}]{Taie:2010uo}%
  \BibitemOpen
  \bibfield  {author} {\bibinfo {author} {\bibfnamefont {S.}~\bibnamefont
  {Taie}}, \bibinfo {author} {\bibfnamefont {Y.}~\bibnamefont {Takasu}},
  \bibinfo {author} {\bibfnamefont {S.}~\bibnamefont {Sugawa}}, \bibinfo
  {author} {\bibfnamefont {R.}~\bibnamefont {Yamazaki}}, \bibinfo {author}
  {\bibfnamefont {T.}~\bibnamefont {Tsujimoto}}, \bibinfo {author}
  {\bibfnamefont {R.}~\bibnamefont {Murakami}}, \ and\ \bibinfo {author}
  {\bibfnamefont {Y.}~\bibnamefont {Takahashi}},\ }\href@noop {} {\bibfield
  {journal} {\bibinfo  {journal} {Physical Review Letters}\ }\textbf {\bibinfo
  {volume} {105}},\ \bibinfo {pages} {190401} (\bibinfo {year}
  {2010})}\BibitemShut {NoStop}%
\bibitem [{\citenamefont {Cohen-Tannoudji}\ \emph {et~al.}(1992)\citenamefont
  {Cohen-Tannoudji}, \citenamefont {Dupont-Roc},\ and\ \citenamefont
  {GRYNBERG}}]{CohenTannoudji:1992wp}%
  \BibitemOpen
  \bibfield  {author} {\bibinfo {author} {\bibfnamefont {C.}~\bibnamefont
  {Cohen-Tannoudji}}, \bibinfo {author} {\bibfnamefont {J.}~\bibnamefont
  {Dupont-Roc}}, \ and\ \bibinfo {author} {\bibfnamefont {G.}~\bibnamefont
  {GRYNBERG}},\ }\href@noop {} {\emph {\bibinfo {title} {{Atom-Photon
  Interactions}}}}\ (\bibinfo  {publisher} {Wiley},\ \bibinfo {address} {New
  York},\ \bibinfo {year} {1992})\BibitemShut {NoStop}%
\bibitem [{\citenamefont {Burdick}\ \emph {et~al.}(2016)\citenamefont
  {Burdick}, \citenamefont {Tang},\ and\ \citenamefont
  {Lev}}]{2016arXiv160503211B}%
  \BibitemOpen
  \bibfield  {author} {\bibinfo {author} {\bibfnamefont {N.~Q.}\ \bibnamefont
  {Burdick}}, \bibinfo {author} {\bibfnamefont {Y.}~\bibnamefont {Tang}}, \
  and\ \bibinfo {author} {\bibfnamefont {B.~L.}\ \bibnamefont {Lev}},\
  }\href@noop {} {\bibfield  {journal} {\bibinfo  {journal} {Physical Review
  X}\ }\textbf {\bibinfo {volume} {6}} (\bibinfo {year} {2016})}\BibitemShut
  {NoStop}%
\bibitem{sup} See Supplemental Material at [URL will be inserted by publisher].

\bibitem [{\citenamefont {Cui}\ \emph {et~al.}(2013{\natexlab{b}})\citenamefont
  {Cui}, \citenamefont {Lian}, \citenamefont {Ho}, \citenamefont {Lev},\ and\
  \citenamefont {Zhai}}]{Cui:2013ki}%
  \BibitemOpen
  \bibfield  {author} {\bibinfo {author} {\bibfnamefont {X.}~\bibnamefont
  {Cui}}, \bibinfo {author} {\bibfnamefont {B.}~\bibnamefont {Lian}}, \bibinfo
  {author} {\bibfnamefont {T.-L.}\ \bibnamefont {Ho}}, \bibinfo {author}
  {\bibfnamefont {B.~L.}\ \bibnamefont {Lev}}, \ and\ \bibinfo {author}
  {\bibfnamefont {H.}~\bibnamefont {Zhai}},\ }\href@noop {} {\bibfield
  {journal} {\bibinfo  {journal} {Physical Review A}\ }\textbf {\bibinfo
  {volume} {88}},\ \bibinfo {pages} {011601} (\bibinfo {year}
  {2013}{\natexlab{b}})}\BibitemShut {NoStop}%

\end{thebibliography}

%

\end{document}